\documentclass[
 reprint,
amsmath,amssymb,
aps,
pra,
]{revtex4-2}
\usepackage{hyperref}
\usepackage{mathtools}
\usepackage{graphicx}% Include figure files
\usepackage{dcolumn}% Align table columns on decimal point
\usepackage{bm}% bold math
\usepackage{multirow} % multiple rows

\begin{document}

\title{Twisted Polarization Domains and Their Dynamics}

\author{Apostolos Brimis$^{1,3,}$} 
\email[Corresponding author: ]{abrimis@physics.uoc.gr}
\author{Konstantinos G. Makris$^{1,3}$}
\author{Dimitris G. Papazoglou$^{2,3}$}

\affiliation{$^1$ITCP, Department of Physics, University of Crete, 70013, Heraklion, Greece}
\affiliation{$^2$Department of Materials Science and Technology, University of Crete, P.O. Box 2208, 70013, Heraklion, Greece}
\affiliation{$^3$Institute of Electronic Structure and Laser, Foundation for Research and Technology-Hellas (FORTH), P.O. Box 1527, 71110, Heraklion, Greece}

%\date{\today}

\begin{abstract}
We provide a theoretical investigation of optical Poincaré beams that exhibit interwinding chiral polarized domains upon propagation. We employ both analytical and numerical methods in order to investigate this phenomenon. Specifically, we introduce the theoretical framework that elucidates the formation and spiraling behavior of chiral polarized domains of light. Additionally, we define dynamic quantities that help us understand and quantify the angular motion of these domains. We apply this method to cylindrically symmetric optical beams, thereby unveiling their distinct radial and longitudinal propagation dynamics.
\end{abstract}

\maketitle

\section{Introduction}
In recent years, the controlled manipulation of light has garnered significant attention in the field of optics \cite{Forbes2021}. Structured light has a wide range of modern applications across several fields, including physics \cite{He2022}, biology \cite{Angelo2018}, and engineering \cite{Liberadzki2018}. In optics, structured light has been utilized to enhance imaging resolution, allowing for better observation of biological samples at the cellular and subcellular levels \cite{Wu2018}. Additionally, the manipulation of light's properties has opened up new avenues for optical communication \cite{Wang2012, Bozinovic2013} since information can be also encoded by light structuring. Likewise, it can be advantageous for a broad range of applications \cite{Oldenbourg1996,Rosales-Guzman2018,Wang2020,He2021} to independently, or collectively, engineer the phase, the amplitude, and the polarization state of  light. 

Polarization is a fundamental property of light, that refers to the direction of the electric field oscillation on a plane, transverse to the propagation direction. Depending on the trajectory of the electric field vector, light polarization can be linear, circular, or elliptical. In particular, circularly polarized light carries {s}pin {a}ngular {m}omentum (SAM) per photon of magnitude $\pm \hbar$, depending on the handedness of the polarization \cite{Born1999}. Conversely, in the presence of a phase singularity, in the form of an optical vortex, light carries {o}rbital {a}ngular {m}omentum (OAM) per photon \cite{Allen1992}. Unlike SAM, the magnitude of OAM is unbounded {\cite{Fickler}} taking values of $l \hbar$, where $l$ is the topological charge, indicating the number of $2\pi$ phase turns around a circular path. Generation and manipulation of OAM finds use in a wide range of applications, such as particle manipulation \cite{He2021, Paterson2001, Grier2003} and quantum communication \cite{Barreiro2008}. 

In addition, structured optical beams, such as Bessel beams \cite{Durnin1987, Durnin2005}, Bessel-Gauss beams \cite{Gori} and Laguerre-Gaussian beams \cite{Siegman1986,Paufler2019}, have found extensive use in laser-based applications. Bessel and Bessel-Gauss beams exhibit non-diffracting and quasi-nondiffracting properties, allowing for long-distance propagation and minimizing diffraction effects. Laguerre-Gaussian beams, despite being diffracting, are valuable due to their support in common laser cavities. Another type of structured light, the Airy beam \cite{Siviloglou2007,Siviloglou2007exp}, demonstrates a unique feature of intensity following a parabolic trajectory as it propagates, known as "acceleration" \cite{Siviloglou2007,Siviloglou2007exp}. By applying the Airy distribution along the radial coordinate of a cylindrically symmetric beam, the so-called ring-Airy beams are formed~\cite{Efremidis2010,Papazoglou2011}. These beams, exhibit
abrupt self-focusing and have been widely applied in material processing and laser-based applications~\cite{Efremidis2019}. Recent research has focused on combining circularly symmetric optical beams carrying OAM \cite{Schechner1996,Bekshaev2006,Bekshaev2006_2}, resulting in varied propagation dynamics such as angular rotation \cite{McGloin2003,Vasilyeu2009,Rop2012}, angular acceleration \cite{Schulze2015,Webster2017}, and tornado-like motion \cite{Brimis2020, Mansour2022,Brimis2023}.

On the other hand, in the case of so-called Poincaré beams \cite{Beckley2010}, the polarization of light is structured, resulting in complex polarization distributions. Poincaré beams can be generated by combining two co-propagating or counter-propagating vortex beams {in an orthogonal polarization} basis \cite{Lopez-Mago2019,Holmes2019,Arora2019,Ruchi2020,Droop2021}. The parameters that determine their properties include the spatial distribution of their field components, their relative weighted amplitude, their relative phase, and the polarization basis used for analysis. As a result, the complex polarization properties of Poincaré beams can be utilized to generate electromagnetic fields with customized properties \cite{Milione2011,Galvez2012,Alpmann2017}. These beams carry both {SAM} and OAM and are characterized by a spatial variation of polarization, including polarization singularities \cite{Dennis2009,Shvedov2015} where the polarization state of light becomes undefined. Additionally, they can induce optical forces on submicron particles \cite{Wang3} and exhibit intriguing propagation dynamics such as optical activity in free space \cite{Sanchez-Lopez2019,Li2020}. Moreover, polarization oscillating beams \cite{Li2018} and angularly accelerating polarization structures \cite{Singh2021} have also been reported.

Although the concept of rotating polarization domains has been reported previously \cite{Singh2021,Krasnoshchekov2017,Krasnoshchekov2019}, there is a lack of the theoretical understanding and systematic analysis of the formation and evolution of such domains. In this manuscript, we examine Poincaré beams with chiral polarized domains, that spiral along the propagation direction.  We provide a theoretical basis to explain the formation of such domains and analyze their complex transverse and longitudinal evolution. In particular, we demonstrate that by superimposing two cylindrically symmetric vortex beams with {different}  OAM  {states $(l_x \neq l_y)$} in the cartesian basis of polarization, angularly rotating and accelerating chiral polarized domains of light can be generated. By applying this method to different families of optical beams, we establish a direct relationship between the number of domains and the associated topological charges of the polarization components. 

%%%%%%%%%%%%%%%%%%%%%%%%%%%%%%%%%%%%%%%%%%%%%%%%%%%%%%%%%%%%%%%%%%%%%

\section{\label{section1} Chiral polarized domains of optical beams}

Let us start by considering the paraxial wave equation of diffraction, which is: 
\begin{equation}
     {i\frac{\partial U}{\partial z} + \frac{1}{2k}\nabla_{\bot}^2 U  = 0}
    \label{eq:paraxial}    
\end{equation}

where $\nabla_{\bot}^2$ denotes the transverse part of the Laplacian,  $z$ is the propagation distance, $U$ is the electric field's envelope and $k$ is the free space wavenumber. 

Working on a cartesian basis of polarization, we assume a superposition of horizontal and vertical polarization components defined as:
\begin{equation}
    \label{eq:eq1}   
    U(r,\theta,z)= \frac{1}{\sqrt{2}} [U_x(r,\theta,z) \hat{\textbf{x}} + U_y (r,\theta,z)  \hat{\textbf{y}}]
\end{equation}
where  $\hat{\textbf{x}}, \hat{\textbf{y}}$ denote the unity vectors of horizontal and vertical polarization, respectively while $U_x, U_y$ are the complex amplitudes of horizontal and vertical components respectively. In this manuscript, we will focus our attention on different types of circularly symmetric vortex beams, like Bessel beams \cite{Durnin1987}, Laguerre-Gaussian beams \cite{Paufler2019}, and ring-Airy beams \cite{Efremidis2010,Papazoglou2011} with vortex \cite{Jiang2012}. In this context, due to the cylindrical symmetry, we can separate the spatial $(r,z)$  and the azimuthal $\theta$ dependence, as $U(r,\theta,z)=U(r,z) U(\theta)$. By applying the aforementioned condition we can rewrite the superposition as:
\begin{equation}
\label{eq:eq3}   
{U(r,\theta,z)= \frac{1}{\sqrt{2}}[A_x \hat{\textbf{x}} + A_ye^{i[\Delta l - \Delta\Phi\theta]}  \hat{\textbf{y}}]e^{i[\Phi_x + l_x\theta]}}
\end{equation}
where $A_x, A_y$ are the magnitude of the horizontal and vertical components, $\Delta \Phi=\Phi_x(r,z) - \Phi_y(r,z)$, where $\Phi_x(r,z)$, $\Phi_y(r,z)$, are the corresponding reduced spatial phases, and $\Delta l=l_y - l_x$, where $l_j, (j:x,y)$ is the topological charge of optical vortices represented by $e^{il_j\theta}$ phase terms.

In general, the polarization of such superimposed fields is nonuniform. A quantity that can be used to identify the regions where the polarization is right and left-handed is the ellipticity angle $\chi$ of the polarization ellipse \cite{Born1999}, which is defined as: 
\begin{equation}
    sin(2 \chi)=\frac{2 A_x A_y sin(\delta)}{A^2_x+A^2_y}
    \label{eq:eq6}    
\end{equation}
where $\delta=\Delta \Phi -\Delta l \theta$. Depending on their parameters various cases can be examined. For instance, if we assume that the vortices carry same topological charge ($l_x=l_y$) or that no vortices exist ($l_x, l_y=0$), and the field’s components are not identical, the ellipticity angle is modulated within circular concentric left and right-handed polarized regions. The chirality of these regions changes along the propagation direction depending on their phase difference $\Delta \Phi$.
On the other hand, by proper selection of the topological charges, azimuthally modulated domains of chiral polarization can be formed. More specifically, when the two components carry OAM of opposite handedness ($l_x \cdot l_y < 0$), it is straightforward to show that $N=|l_x|+|l_y|$ chiral polarized domains that are separated by lines of linear polarization (L-lines) \cite{Nye_book,Freund2002b}, are formed. {On the other hand, when they carry OAM of same handedness ($l_x \cdot l_y > 0$), $N=||l_x|-|l_y||$ chiral polarized domains are formed.} {For both cases, t}he  whole distribution  is oriented at an angle $\theta$, defined by: 
\begin{equation}
    \label{eq:eq4}   
    \theta = \frac{1}{N} {\Delta \Phi({r,z}}).  
\end{equation}

As we observe, the angle $\theta$ is proportional to $\Delta \Phi$ and inversely proportional to $N$, keeping in mind that in some cases $\Delta \Phi$ can also{depend} on the topological charges {as in the case of Laguerre-Gaussian beams (see section \ref{LG})}. By calculating the first and the second derivative with respect to the propagation distance $z$, we derive  the angular velocity and the angular acceleration \cite{Brimis2020,Mansour2022} as:  
\begin{equation}
    \label{eq:eq5}   
    \upsilon  \equiv \frac{{\partial \theta }}{{\partial z}}= \frac{1}{N}\frac{{\partial\Delta \Phi }}{{\partial z}}, \qquad
    \gamma \equiv \frac{\partial \upsilon}{\partial z} = \frac{1}{N}\frac{{{\partial^2}\Delta \Phi }}{{\partial{z^2}}}    
\end{equation}

Eq. (\ref{eq:eq4}) and Eqs. (\ref{eq:eq5}) describe the propagation dynamics of the spiraling  polarization domains. Following a similar approach, equations identical to Eqs. (\ref{eq:eq4}),(\ref{eq:eq5}), were shown in \cite{Mansour2022} to describe the propagation dynamics of high-intensity lobes, in the case of scalar superposition of opposite-handedness OAM beams. Interestingly, there is a one-to-one analogy between the two schemes. More specifically, for the superposition of scalar fields with opposite OAM handedness, the field intensity is modulated and rotates along propagation \cite{McGloin2003,Vasilyeu2009,Rop2012,Schulze2015,Webster2017,Brimis2020,Mansour2022,Kotlyar1997,Kotlyar2007,Wu2021,Jiang2022,Zhang2022,Xu2023}, while in our case where  polarization components carrying opposite OAM handedness are superimposed, the field polarization is modulated with polarization domains{that} rotate along propagation. Now that we have presented the theoretical framework of our study, we will apply it, in the next paragraphs, to different families of optical beams, namely Bessel beams, Laguerre-Gaussian beams and ring-Airy vortex beams.

%%%%%%%%%%%%%%%%%%%%%%%%%%%%%%%%%%%%%%%%%%%%%%%%%%%%%%%%%%%%%%%%%%%%%%%%%
%%%%%%%%%%%%%%%%%%%%%%%%%%%%%%%%%%%%%%%%%%%%%%%%%%%%%%%%%%%%%%%%%%%%%%%%%%%

\section{\label{BESSEL}Bessel beams}

As a first example, let us consider the case of an  {ideal non-diffracting}  Bessel beam \cite{Durnin1987,Durnin2005}. The two perpendicular components are described as:
\begin{equation}
    \label{eq:eq7}   
    U_j(r,\theta,z)=J_{l_j}(k_{T_{j}}{r}){e^{i{l_j}{\theta}}}{e^{ik_{z_{j}}z}}
\end{equation}
where $J_{l_j}(\cdot)$ ($j:x,y$) is the $l_j$-order Bessel function and $r$ is the radial coordinate. For each beam, $k_{T_j}, k_{z_j}$ ($j=x,y$), refers to the transverse and the longitudinal components of the wavenumber for which $k_{z_j} = \sqrt{{k_0}^2-{k_{T_j}}^2}$, and $k_0= 2 \pi/ \lambda$ is the vacuum wavenumber. In this case, $\Delta \Phi=(k_{z_x}-k_{z_y})z$ varies linearly with the propagation distance and Eq. (\ref{eq:eq4}) is now expressed as:
\begin{equation}
    \label{eq:eq8}   
    \theta = \frac{k_{z_x}-k_{z_y}}{N} z.  
\end{equation}
This linear dependence, as it can be seen from Eq.  (\ref{eq:eq5}), leads to a constant angular velocity for such domains and no angular acceleration occurs. 
\begin{equation}
    \label{eq:eq9}   
    \upsilon = \frac{k_{z_x}-k_{z_y}}{N}= const., \qquad
    \gamma  = 0   
\end{equation}
\begin{table}[!b]
%\color{red}
\centering
\caption{{Parameters of Bessel beams.}}
\label{Tab:bb table}
\begin{tabular}{ccccc}
\hline
\multirow{2}{*}{Bessel beam}       & $k_T$          & \multirow{2}{*}{$l$}             & $\lambda$ &  \\
                           & $(mm^{-1})$          &                & $(nm)$ &   \\ \hline
$\hat{x}$                   & 27.5                   &  (-1,-1,-2,{1})            & 800  &    \\
$\hat{y}$                   & 31.4                  &   (1,2,2,{3})         & 800  &    \\ \hline
%27.4889, 31.4158
\end{tabular}
\end{table}

\begin{figure}[t!]
\centering
\includegraphics[width=0.45\textwidth]{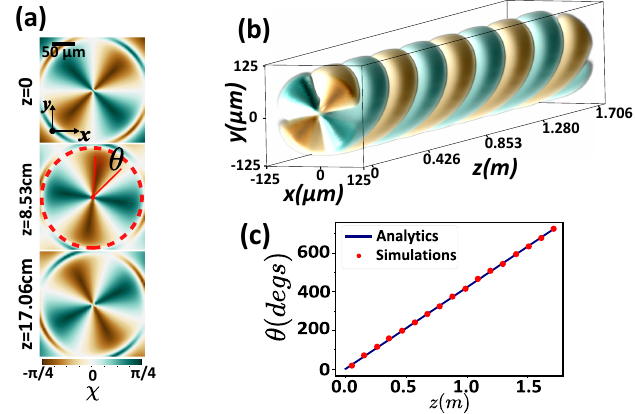}
\caption{Spiraling chiral domains that are formed by Bessel beams with ($l_x=-1$, $l_y=1$). (a) Transverse profiles of ellipticity angle $\chi$ along the propagation. {Positive values of the angle $\chi$ correspond to left-handed polarization, while negative values correspond to right-handed polarization. When $\chi=0$, the polarization is linear.} (b) Three-dimensional illustration of ellipticity angle along propagation for the isolated region enclosed with the red dotted circle of (a). (c) Angular position of the rotating chiral domains as a function of the propagation distance.}
\label{fig:bessel1}
\end{figure}

\begin{figure}[b!]
\centering
\includegraphics[width=0.45\textwidth]{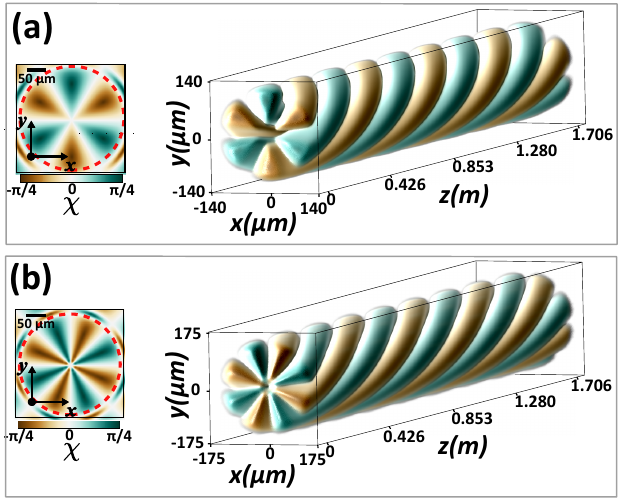}
\caption{Three-dimensional illustration of the spiraling chirality that is produced by Bessel beams. (a)-(b) Three-dimensional representation of spiraling ellipticity angle upon propagation for the green dotted circled region produced using different combinations of topological charges as (a) ($l_x=-1$, $l_y=2$) and (b) ($l_x=-2$, $l_y=2$). The insets in both cases show a typical cross sectional profile of the ellipticity angle.}
\label{fig:bessel2}
\end{figure}

Likewise, we have performed numerical simulations by solving the paraxial wave equation of such a superposition. The Bessel beams' parameters that have been used are presented in {Table} \ref{Tab:bb table}. In Fig. \ref{fig:bessel1} we present the  evolution of the polarization domains that are formed by combining such high-order Bessel beams. In Fig. \ref{fig:bessel1}(a)  transverse images of the ellipticity angle at different propagation distances are shown. The distribution varies from right to left-handed polarization while regions of linear polarization, forming the L-lines, are visualized in white. Due to their non-diffracting nature,  these polarization domains retain their radial shape while they rotate as it can be analytically shown. In Fig. \ref{fig:bessel1}(b) a three-dimensional representation of the ellipticity angle is presented. Clearly, as they propagate these domains rotate, forming spiraling chiral domains. In order to quantify this behavior, we measure the angular position of their maximum  ellipticity value. The simulation data are presented in Fig. \ref{fig:bessel1}(c)  along with the prediction based on the analytical expression of Eq. (\ref{eq:eq8}). It is evident that there is an excellent agreement between theoretical prediction and numerical simulations. Furthermore, as expected (see Eq. (\ref{eq:eq8})) the angular position is a linear function of the propagation distance thus the alternate polarization domains rotate at a constant rate and no angular acceleration occurs.

We now further investigate the behavior of such spiraling chiral polarized domains for different combinations of topological charges. Especially, in Fig. \ref{fig:bessel2} we present the ellipticity angle for  ($l_x=-1$, $l_y=2$) and ($l_x=-2$, $l_y=2$). In Fig. \ref{fig:bessel2}(a) a transverse profile is presented for the case of ($l_x=-1$, $l_y=2$). Clearly, $N=|l_x|+|l_y|=3$ polarization domains  are formed forming spiral polarization domains as they propagate. In the same content, in Fig. \ref{fig:bessel2}(b) a transverse profile is presented for the case of ($l_x=-2$, $l_y=2$). As expected,  $N=4$ spiraling chiral polarized domains are now formed. 

\begin{figure}[t!]
    \centering
    \includegraphics[width=0.45\textwidth]{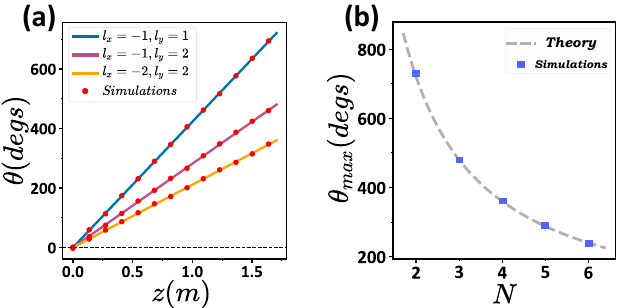}
    \caption{Effect of the topological charge on the evolution of spiraling chiral polarized domains produced by Bessel beams superposition. (a) Angular position as a function of propagation distance for all the presented combinations of topological charges. (b) Maximum angular position as a function of $N$-domains at a specific propagation distance ($z=1.706 m$).}
    \label{fig:bessel3}
\end{figure}

The effect of the topological charge is presented on the angular position as a function of propagation distance in Fig. \ref{fig:bessel3}(a). As predicted by the theory, the topological charge affects only the rate of the linear dependence. Furthermore, Fig. \ref{fig:bessel3}(b) shows a very good agreement of the numerical simulations to a $1/N$ power law as predicted by Eq. (\ref{eq:eq8}). 
Based on Eq. (\ref{eq:eq8}), different choices of $l_x, l_y$ can result in the same number of chiral domains with the same angular behavior. In addition, the sign of the rotation is determined solely by the selection of the component with the positive/negative topological charge and the difference between the longitudinal projection of the wavenumber $k_z$.

\begin{figure}[t!]
    \centering
    \includegraphics[width=0.45\textwidth]{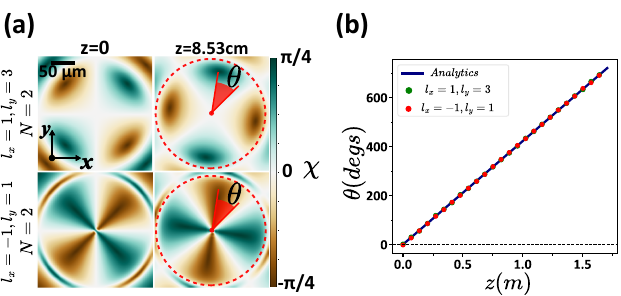}
    \caption{{Effect of the topological charges' sign on the formation and evolution of chiral polarized domains produced by Bessel beams superposition. (a) Transverse profiles of ellipticity angle $\chi$ along the propagation. Positive values of the angle $\chi$ correspond to left-handed polarization, while negative values correspond to right-handed polarization. When $\chi=0$, the polarization is linear. (b)Angular position as a function of propagation distance for the two presented cases.}}
    \label{fig:bessel4}
\end{figure}

{Finally let's investigate the effect of the sign of the topological charges on the formation and dynamics of the chiral polarized domains. As analyzed in the previous section, the number of chiral polarized domains depends on both the magnitude and the sign of the topological charges. In Fig. \ref{fig:bessel4}, we present two cases, one for $l_x \cdot l_y>0$  and one for $l_x \cdot l_y<0$. Specifically, in Fig. \ref{fig:bessel4}(a), we show transverse profiles of the ellipticity angle at different propagation distances for the case of $l_x=1$, $l_y=3$ (upper row) and $l_x=-1$, $l_y=1$ (lower row). As we observe, although the number of chiral polarized domains in both cases is the same ($N = 2$), the chiral domains are quite different in shape. This was expected since the shape strongly depends on the overlapping between the two field components. Additionally, in Fig. \ref{fig:bessel4}(b), we present the angular position along the propagation for both cases. It is evident that their dynamics are almost identical, a result consistent with Eq. (\ref{eq:eq8}), revealing that the dynamics only depend on the number of chiral polarized domains and not on the sign of the topological charges, for the case of Bessel beams.  }

{Note that up until now, we have presented the ideal non-diffracting Bessel beams, which are not be experimentally  realizable due to the infinite energy required. Their experimentally  realizable counterpart  are the well-known Bessel-Gauss beams \cite{Gori}, which are diffracting solutions of the paraxial wave equation. However, with a proper choice of the width of the Gaussian envelope, quasi-nondiffracting chiral polarized domains can be achieved. In this case, the angular position linearly increases with the propagation distance, similar to that of the ideal non-diffracting Bessel beams, but for smaller propagation distances.}

%%%%%%%%%%%%%%%%%%%%%%%%%%%%%%%%%%%%%%%%%%%%%%%%%%%%%%%%%%%%%%%%%%%%%%%%%
%%%%%%%%%%%%%%%%%%%%%%%%%%%%%%%%%%%%%%%%%%%%%%%%%%%%%%%%%%%%%%%%%%%%%%%%%%%

\section{\label{LG}Laguerre-Gaussian beams} 
%\textbf{DP: continue from here}

Examining the case of Laguerre-Gaussian beams (LG) \cite{Siegman1986,Paufler2019} is the next step. Each component's analytic expression along the propagation is given by the expression:

\begin{equation}
\label{eq:eq10}   
\begin{split}
U_j(r,\theta,z) = \frac{A_j}{w_j(z)} \bigg(\frac{\sqrt{2}r} 
{w_j(z)}\bigg)^ {|l_j|} L_{p_j}^{|l_j|} \bigg (\frac{2 {r}^2}{w_j(z)^2}\bigg)  \\
\exp\bigg[ik\frac{{r} ^2}{2R_j(z)} \bigg]  \exp[-i\zeta_j(z)] \\   \exp[il_j \theta]
\exp\bigg[-\frac{{r}^2}{w_j(z)^2}\bigg], j:x,y
\end{split}
\end{equation}

where,

\begin{equation}
\label{eq:eq111} 
\begin{split}
A_j= \sqrt{{\frac{2 p_j ! }{\pi (\mid l_j\mid+p_j) ! }}} \\
w_j(z)=w_{0_j} \sqrt{1+\bigg(\frac{z}{z_{R_j}}\bigg)^2} \\ 
\notag %-> this is for not numbering
R_j(z)=z\bigg[ 1+\bigg(\frac{z}{z_{R_j}}\bigg)^2\bigg] \\
\zeta_j(z)=(2p_j+|l_j|+1)\arctan(\frac{z}{z_{R_j}}).
\end{split}
\end{equation}

For each polarization component ($j:x,y$), $L_{p_j}^{|l_j|}(\cdot)$ is the associated Laguerre polynomial, $A_j$ is a normalization constant, $w_j(z)$ is the spot size at distance $z$, $z_{R_j}$ is the Rayleigh length, $R_j(z)$ is the radius of curvature,  $\zeta_j(z)$ is the Gouy phase,  $l_j$  is the topological charge of the vortex and $p_j$ is a radial index. Now there are two methods to construct beams that exhibit angularly rotating chiral polarized domains, and both of them are based on superposition of LG beams with opposite OAM handedness. The first approach is to combine components with distinct widths $w_{0_j}$ and, consequently, Rayleigh lengths $z_{R_j}$. This will result in a Gouy phase difference and, therefore, an evolving nonzero reduced phase difference along the propagation $z$. Combining components with identical widths but distinct radial indices $p_{j}$ is the second approach. This leads to distinct Gouy phases $\zeta_{j}(z)$ resulting in a nonzero reduced phase difference. Here, for the sake of clarity, we present the second case. Therefore, the reduced phase difference can be expressed as $\Delta \Phi=[2(p_y-p_x)+(|l_y|-|l_x|)]arctan(z/z_R)$, and Eq. (\ref{eq:eq4}) is now written as: 

\begin{equation}
\label{eq:eq12}   
\theta = \frac{ 2 \Delta p + \Delta |l|}{N} arctan(z/z_R)  
\end{equation}
where $\Delta p =p_y-p_x$ and $\Delta |l| =|l_y|-|l_x|$. If we apply to this case the general Eqs. (\ref{eq:eq5}), the angular velocity and angular acceleration can be rewritten as:  
\begin{equation}
\label{eq:eq13}   
%\begin{split}
\upsilon  = \frac{2\Delta p+\Delta |l|}{N}\frac{z_R }{z^2+z_R^2}, %\\
\gamma  = -2\frac{2\Delta p +\Delta |l|}{N}\frac{z z_R}{\Big(z^2+{z_R}^2\Big)^2} 
%\end{split}
\end{equation}

\begin{figure}[t!]
\centering
\includegraphics[width=0.45\textwidth]{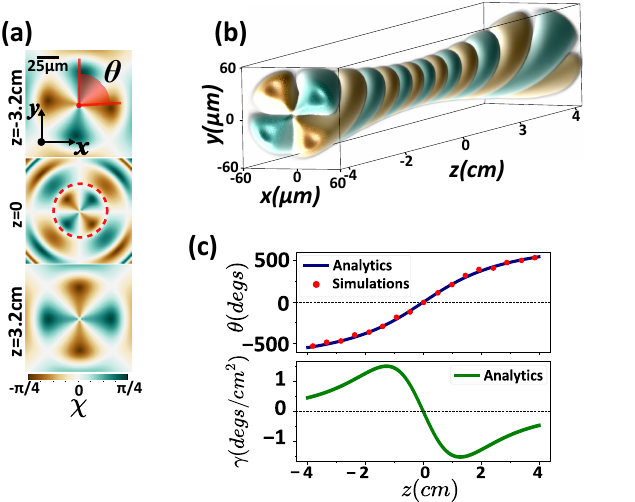}
\caption{Spiraling chiral domains that are formed by Laguerre-Gaussian beams with ($l_x=-1$, $l_y=1$). (a) Transverse profiles of ellipticity angle $\chi$ along the propagation.{Positive values of the angle $\chi$ correspond to left-handed polarization, while negative values correspond to right-handed polarization. When $\chi=0$, the polarization is linear.} (b) Three-dimensional illustration of ellipticity angle upon propagation for the prescribed region enclosed by the red dotted circle in (a). (c) Angular position (upper subplot) and angular acceleration (lower subplot) of the rotating chiral domains as a function of propagation distance. }
\label{fig:LG1}
\end{figure}

Such a nonlinear dependence of $\theta$ results in a variable angular velocity $\upsilon$ of the ellipticity angle, as acceleration/deceleration $\gamma$ is now occurring. By examining the above equations for $|z|\gg z_R$,  $\lim_{|z| \to \infty} atan(\frac{z}{z_R}) = \frac{\pi}{2}$, we can determine the maximum angular position of the ellipticity angle, which is $\theta_{max}=\frac{(2\Delta p +\Delta |l|) \pi}{2N}$, whereas the velocity and angular acceleration tend to zero. Additionally, the rapid rotation occurs at the beam's waist ($z=0$). 

\vspace{0.1cm}
\begin{table}[!b]
\centering
\label{Tab:parameters}
\caption{{Parameters of Laguerre-Gaussian beams.}}
\label{Tab:lg table}
\begin{tabular}{cccccc}
\hline
\multirow{2}{*}{LG beam}       & $w_0$          & \multirow{2}{*}{$l$}       & \multirow{2}{*}{$p$}        & $\lambda$ &  \\
                           & $(\mu\text{m})$          &          &           & $(nm)$ &   \\ \hline
$\hat{x}$                   & 74.9                  &  (-1,-2,-2)  & 1            & 800  &    \\
$\hat{y}$                   & 74.9                 & (1,1,2) &  10             & 800  &    \\ \hline
\end{tabular}
\end{table}

Furthermore, we numerically solved the paraxial wave equation and compared our results to the analytic predictions. Table \ref{Tab:lg table} presents the parameters of the beams that were used. Fig. \ref{fig:LG1} shows the spiraling chiral polarized domains created by superimposing LG beams. Specifically, Fig. \ref{fig:LG1} (a) depicts transverse images of the ellipticity angle before and after the beam waist. $N=2$ chiral polarized domains were separated by L-lines to form the pattern. Additionally, as the domains propagate, they grow radially until they reach their smallest size at the beam's waist plane. In Fig. \ref{fig:LG1}(b) a three-dimensional representation of the ellipticity angle is shown. The arms of chiral polarized domains are intertwined to form a spiral with variable pitch and radius, whereas the rotation around the waist plane is more rapid. Due to diffraction, the chirality follows the evolution of the width, and the central chiral polarized domains grow in size far from the waist plane, as is evident. We have determined the angular position along the propagation axis in order to quantify the angular position of such rotating domains. The upper graph in  Fig. \ref{fig:LG1}(c) illustrates the angular position as a function of propagation distance for both simulations and the analytical results. As we can see, we have excellent agreement. Due to the nonlinear dependence of the angular position regarding the propagation distance, we expect angular acceleration to occur. In the lower graph of Fig. \ref{fig:LG1}(c), the angular acceleration of such a structure is presented. As the beam propagates from negative to positive propagation distances, the chiral polarized domains accelerate reaching their maximum value.

\begin{figure}[t!]
\centering
\includegraphics[width=0.45\textwidth]{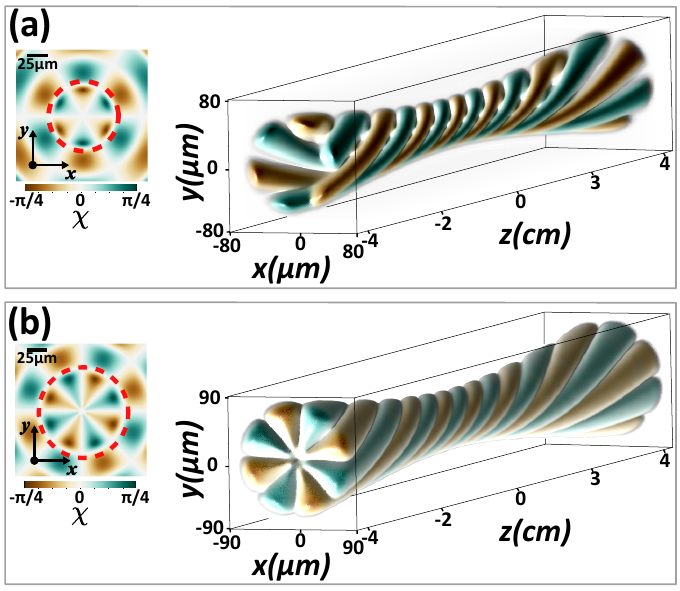}
\caption{Three-dimensional plot of the spiraling chirality produced by Laguerre-Gaussian beam superposition. (a)-(b) Three-dimensional representation of spiraling ellipticity angle along propagation axis for the green dotted circled region produced using different combinations of topological charges as (a) ($l_x=-2$, $l_y=1$) and (b) ($l_x=-2$, $l_y=2$). The insets in both cases show a typical cross sectional profile of the ellipticity angle.}
\label{fig:LG2}
\end{figure}

The effect of topological charges on the formation of chiral polarized domains by LG superposition is depicted in Fig. \ref{fig:LG2}. Specifically, the upper left of Fig. \ref{fig:LG2}(a) illustrates the typical cross-sectional distribution of the ellipticity angle at the waist plane of the beam for the case of $l_x=-2$ and $l_y=1$. It is evident that $N=3$ domains appear in the transverse plane around a closed loop. The dashed circle delimits the region depicted in the three-dimensional illustration of the ellipticity angle. Such chiral domains exhibit a spiraling behavior that is intensified near the beam's waist. In addition, they diminish in size around the focus, as is typical for LG beams. In Fig. \ref{fig:LG2}(b), we investigate the situation where $l_x=-2$ and $l_y=2$. We observe $N=4$ domains that rotate about the axis of the beam as it propagates.

\begin{figure}[t!]
\centering
\includegraphics[width=0.45\textwidth]{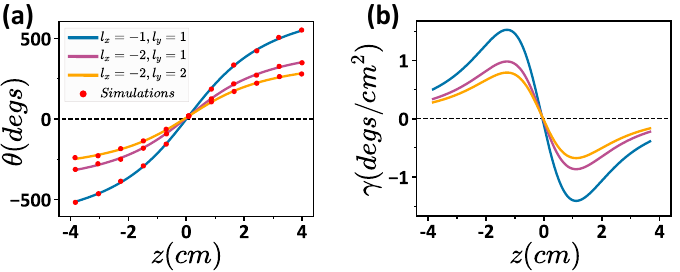}
\caption{Numerical and analytical results for the dynamic quantities of the chiral polarized domains for the presented topological charges configurations. (a) Angular position and (b) angular acceleration of chiral polarized domains for various combinations of topological charges produced by Laguerre-Gaussian beams, as a function of the propagation distance.}
\label{fig:LG3}
\end{figure}

To summarize the effect of the topological charge on the formation and the evolution of chiral polarized domains, we present the angular position and the angular acceleration as a function of propagation distance in Fig. \ref{fig:LG3} for all the presented combinations of topological charges. As we observe in Fig. \ref{fig:LG3}(a), analytics (solid lines) and simulations (red dots) results are in almost perfect agreement. Also, the angular position increases rapidly around $z=0$ while for large $z$, it tends to stabilize. Additionally, in Fig. \ref{fig:LG3}(b) the angular acceleration as described by Eq. (\ref{eq:eq13}) is presented. The acceleration is more profound around the beam waist, while the maximum value depends on the values of the topological charges. 

\begin{figure}[b!]
\centering
\includegraphics[width=0.45\textwidth]{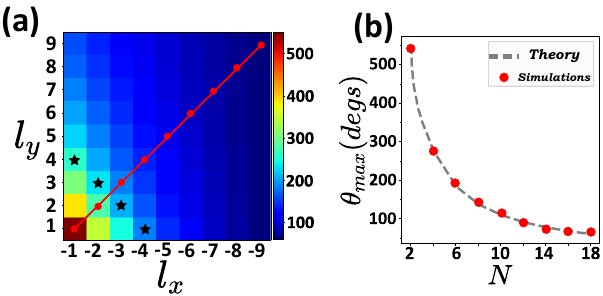}
\caption{Effect of the topological charge on the evolution of spiraling chiral polarized domains produced by Laguerre-Gaussian beams. (a) Maximum angular position as a function of topological charges $l_x, l_y$ at $z=4cm$. The red dots and line, show the special case of $|l_x|=|l_y|$ whereas the black stars show different combinations that produce the same chiral domains $N$. (b) Maximum anglular position as a function of $N=|l_x|+|l_y|$ at a specific propagating distance ($z=4cm$) for the special case where $|l_x|=|l_y|$. The data follow the 1/$N$ power law.}
\label{fig:LG4}
\end{figure}

In contrast to Bessel beams, the relationship between the maximum angle and the topological charges is more complex now. In Eq. (\ref{eq:eq12}), topological charges appear on both $N$ and $\Delta|l|$. For instance, for a nonzero radial difference $\Delta p \neq 0$, stationary chiral polarized domains can be generated when  $\Delta p= -\Delta |l| /2$. Intriguingly, although various configurations of topological charges  {, either with $l_x \cdot l_y > 0$ or $l_x \cdot l_y < 0$,} can generate the same number of $N$ domains, they do not exhibit the same dynamics. This behavior is illustrated in Fig. \ref{fig:LG4}. In particular, Fig. \ref{fig:LG4}(a) depicts the maximum rotation for various combinations of topological charges at a given propagation distance and $\Delta p$. All combinations of $l_x, l_y$ producing the same number of chiral polarized domains ($N=5$) are denoted by black stars. However, their maximal angular positions are distinct. The same holds true for all dynamic quantities, including angular velocity and acceleration.  Moreover, if the topological charges of the two components are swapped, their dynamics will differ. As predicted by Eq. (\ref{eq:eq12}), these results indicate that the power law $1/N$ does not hold for arbitrary parameters of topological charge for LG beams. In addition, red dots illustrate the special case of $|l_x|=|l_y|$ in which the topological charges only appear in $N$ since $\Delta|l|=0$. The power law $1/N$ is now applicable and is presented in Fig. \ref{fig:LG4}(b), while numerical simulations confirm its validity. {As previously mentioned, LG beams have an additional degree of freedom represented by their radial index $p$, allowing for a wide range of possible dynamics by tuning combinations of $p$ and $l$. However, in our study, we focused solely on the influence of the topological charge as it directly relates to the number of chiral polarized domains $N$. Additionally, as indicated in Eq. (\ref{eq:eq12}) and Eqs. (\ref{eq:eq13}), the dependence of the topological charge  is more complex compared to the radial index.}

%%%%%%%%%%%%%%%%%%%%%%%%%%%%%%%%%%%%%%%%%%%%%%%%%%%%%%%%%%%%%%%%%%%%%%%%%
%%%%%%%%%%%%%%%%%%%%%%%%%%%%%%%%%%%%%%%%%%%%%%%%%%%%%%%%%%%%%%%%%%%%%%%%%%%

\section{\label{RA}Ring-Airy Vortex beams}

Since the mechanism responsible for the angular rotation of chiral polarized domains has been clarified, we will apply our method to ring-Airy vortex beams. The scalar superposition of two ring-Airy vortex beams carrying OAM of opposite  sign, known as Tornado Waves (ToWs) \cite{Brimis2020, Mansour2022}, leads to the formation of  high-intensity lobes that accelerate along the radial and angular dimensions as they propagate. The parabolic trajectory of the lobes indicates radial acceleration, whereas the nonlinearly varying phase difference \cite{Mansour2022} is related to angular acceleration. In the same context as before, the initial state of the vertical and horizontal polarization components is defined as follows:

\begin{equation}
\label{eq:14}   
    U_j(r,\theta,0)= Ai({\rho_j}) {e^{a{\rho_j}}} e^{i l_j \theta }
\end{equation}

where $Ai(\cdot)$ ($j:x,y$) is the Airy function, $r$ is the radial coordinate, ${\rho_j} = \left( {{r_{0_j}} - r} \right)/{w_{0_j}}$, $r_{0_j}$ is the radius and $w_{0_j}$ the width parameters of the primary ring \cite{Papazoglou2011,Panagiotopoulos2013} of the ring-Airy beam,  $a$ is an apodization factor and $l_j$ is the topological charge. Unlike the two previous cases, there are no closed form expressions for the complex field's envelope distribution along the propagation axis, so we will present only numerical simulation results by solving the paraxial wave equation Eq. (\ref{eq:paraxial}) using the beam parameters shown in Table \ref{Tab:ra table}.

\begin{figure}[t!]
\centering
\includegraphics[width=0.45\textwidth]{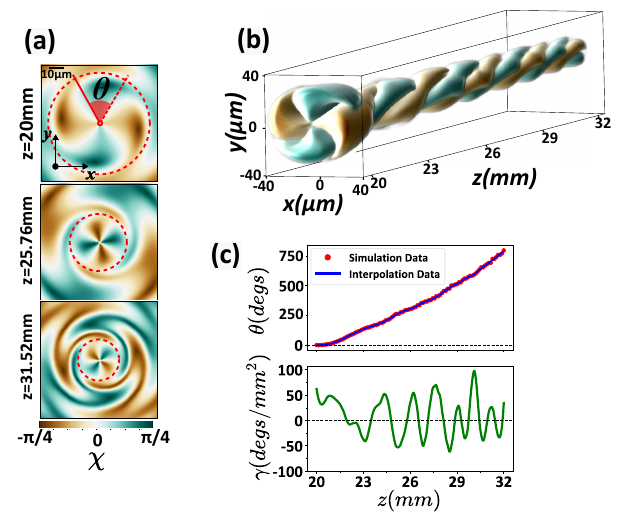}
\caption{Spiraling chiral domains that are formed by superimposing ring-Airy vortex beams with ($l_x=-1$, $l_y=1$). (a) Transverse profiles of ellipticity angle $\chi$ along the propagation.{Positive values of the angle $\chi$ correspond to left-handed polarization, while negative values correspond to right-handed polarization. When $\chi=0$, the polarization is linear.} (b) Three-dimensional plot of ellipticity angle upon propagation for the isolated region enclosed with the red dotted circle in (a). (c) Angular position and angular acceleration of the rotating chiral domains as a function of propagation distance. }
\label{fig:ra1}
\end{figure}

\begin{table}[!b]
%\color{blue}
\centering
\caption{{Parameters of ring-Airy vortex beams.}}
\label{Tab:ra table}
\begin{tabular}{cccccc}
\hline
\multirow{2}{*}{ Ring-Airy}       & $r_0$                 & $w_0$           & \multirow{2}{*}{$a$}   &  \multirow{2}{*}{$l$}  & $\lambda$ \\
                           & $(\mu\text{m})$       & $(\mu\text{m})$ &       &  & $(nm)$  \\ \hline
$\hat{x}$                   & 125                  & 25            & 0.04  &  (-1,-1,-2) & 800      \\
$\hat{y}$                   & 125                & 22.5            & 0.04  & (1,2,2) &  800    \\ \hline
\end{tabular}
\end{table}

In Fig. \ref{fig:ra1} spiraling chiral polarized domains are produced by superimposing ring-Airy vortex beams. Fig. \ref{fig:ra1}(a) depicts the transverse profiles of the ellipticity angle at the region of focus. As observed, the central pattern (enclosed by a red dashed line) consists of $N=|l_y|+|l_x|$ domains of the same handedness that are not distinctly separated from the periphery. In addition, the central region decreases in size as it rotates along the propagation direction. This behavior is also illustrated in Fig.\ref{fig:ra1}(b), where a three-dimensional visualization of the ellipticity angle is presented. We demonstrate that $N=2$ chiral polarized domains rotate along the propagation direction, exhibiting decreasing pitch and size. This is clearly a tornado-like behavior. To quantify their dynamics, we track and measure their angular position along propagation in the focal region. In the upper subplot of Fig. \ref{fig:ra1}(c), the angular position versus propagation distance is displayed. We observe a nonlinear increase along the propagation direction that is modulated by small variations. This complex dependency points out the angular acceleration of such domains. In order to accurately estimate it, the numerical data of the angular position were interpolated and their second derivatives were calculated. In the lower subfigure of Fig. \ref{fig:ra1}(c), the estimated angular acceleration is depicted. As indicated by the graph, chiral polarized domains experience acceleration and deceleration in a quasi-periodic manner along propagation. Such a complex behavior is quite similar to that of ToWs \cite{Brimis2020,Mansour2022}.

\begin{figure}[t!]
\centering
\includegraphics[width=0.45\textwidth]{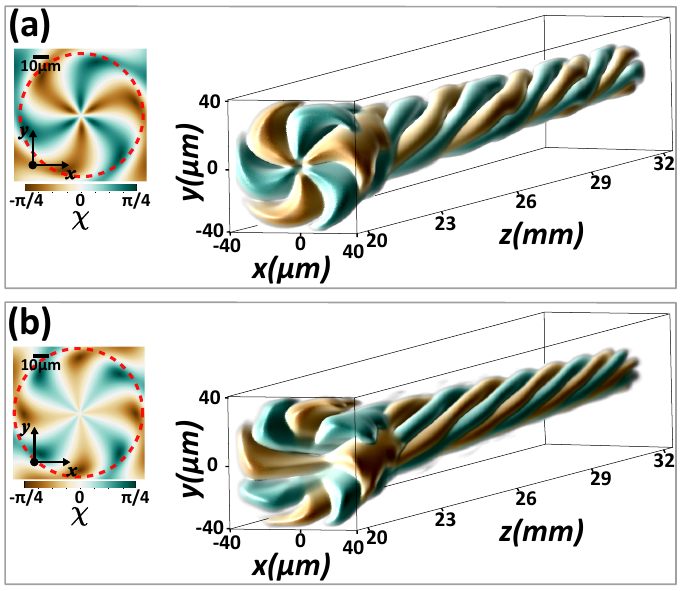}
\caption{Three-dimensional illustration of the spiraling chirality produced by ring-Airy vortex beam superposition. (a)-(b) Three-dimensional representation of spiraling ellipticity angle upon propagation for the green dotted circled region produced using different combinations of topological charges as  (a) ($l_x=-1$, $l_y=2$) and (b) ($l_x=-2$, $l_y=2$). The insets in both cases show a typical cross sectional profile of the ellipticity angle.}
\label{fig:ra2}
\end{figure}

In Figs. \ref{fig:ra2}(a) and (b), we show the ellipticity angle in three dimensions for the cases ($l_x=-1$, $l_y=2$) and ($l_x=-2$, $l_y=1$), respectively. These complex domains consist of $N=3$ and $N=4$ domains with the same polarization handedness. Moreover, in both instances, the structure rotates along propagation with a decreasing pitch and radius, similar to the simplest case of $l_x=-1$ and $l_y=1$.  

\begin{figure}[b!]
\centering
\includegraphics[width=0.45\textwidth]{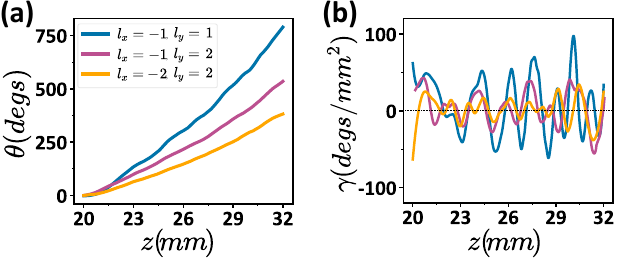}
\caption{Numerical simulation results for the dynamic quantities of the chiral polarized
domains for the presented topological charges configurations. (a) Angular position
and (b) angular acceleration of chiral polarized domains for various combinations of
topological charges produced by ring-Airy vortex beams.}
\label{fig:ra3}
\end{figure}

In Fig. \ref{fig:ra3}, the angular velocity and acceleration are depicted. Fig. \ref{fig:ra3}(a) and (b) present, the angular position and the estimated angular acceleration, respectively. In each of the depicted cases, the angular position increases nonlinearly with small quasi-periodic variations along propagation. In accordance with the general expression of Eq. (\ref{eq:eq4}), the maximum rotation at a particular propagation distance decreases as $N$ increases. In addition, the angular acceleration exhibits oscillatory behavior as it reaches high values upon propagation. As $N$ increases, acceleration and deceleration appear quasi-periodically, while the maximum and minimum values decrease.

\begin{figure}[t!]
\centering
\includegraphics[width=0.45\textwidth]{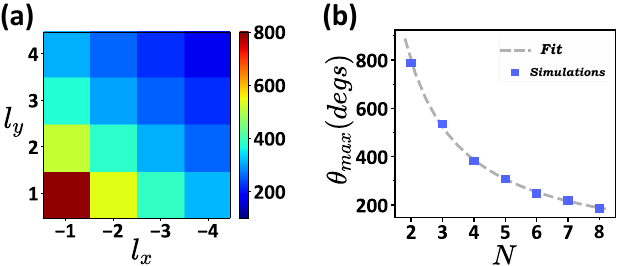}
\caption{Effect of the topological charge on the evolution of spiraling chiral polarized domains produced by ring-Airy vortex beams' superposition. (a) Maximum angular position as a function of topological charges $l_x, l_y$ at $z=32mm$. (b) Maximum angle as a function of $N=|l_x|+|l_y|$ at $z=32mm$. The data follow the 1/$N$ power law.}
\label{fig:ra4}
\end{figure}

To better understand the effect of topological charges, we study the maximum rotation at a given propagation distance for various combinations of topological charges. As shown in Fig. \ref{fig:ra4}(a), in this case, the dynamics do not depend on the values of $l_x$, $l_y$, but rather on the values of $N$. For example, $N=5$ chiral domains may be generated by different topological charge combinations such as ($l_x=-1$, $l_y=4$), ($l_x=-2$, $l_y=3$), ($l_x=-3$, $l_y=2$), and ($l_x=-4$, $l_y=1$). As depicted in the plot, all of them undergo the same dynamics and, consequently, the same maximum angular position.  Moreover, the given set of spatial parameters and topological charges indicate counterclockwise chiral polarized domains, whereas the case of the opposite signs affects only the direction of rotation and not their dynamics. {Additionally, the sign of the vortices does not affect the dynamics of the chiral polarized domains; instead, only $N$ does.} The aforementioned are strong indications that the topological charge of a vortex ring-Airy beam appears only at the vortex phase term and not in any propagation distance-dependent phase term, as in the case of LG beams. Since the topological charge has no effect on the dynamics, we plot the maximal angle as a function of $N$ in Fig. \ref{fig:ra4}(b), regardless of the topological charge configuration used. We clearly observe an excellent match by fitting the data to the $1/N$ power law.

\section{\label{sec:3}Discussion}

As a general conclusion, superimposing vortex beams such as Bessel beams, Laguerre-Gaussian beams, and ring-Airy vortex beams can lead to spiraling Poincaré beams. In the case of Bessel beams, $N$ chiral polarized domains with constant angular velocity are demonstrated, while their radial distribution retains its shape. The angular position of these domains conforms to a simple 1/$N$ power law. In contrast, Laguerre-Gaussian beams exhibit $N$ chiral patterns that experience angular acceleration, which is more pronounced near the beam's waist. The pattern's size decreases and then increases as the beam propagates through the waist. However, the rotation scales as predicted by the  1/$N$ power law only in the special case where $|l_x| = |l_y|$ ($l_x \cdot l_y < 0$). For specific beam parameter selections, stationary patterns can also be generated. The complex behavior of these patterns has been predicted and validated via analytical and numerical results. In the case of ring-Airy vortex beams, $N$ chiral patterns spiral along the propagation direction and display quasi-periodic acceleration and deceleration. Similarly to a tornado, their radius and pitch, diminish. The numerical simulations corroborate the validity of the 1/$N$ power law, which indicates that when a vortex is imprinted in a ring-Airy beam, the topological charge only appears on the vortex term, and no additional topological charge-dependent terms are produced.

\section{Conclusions}
In this study, we systematically study twisted Poincaré beams. We have demonstrated the existence of spiraling chiral polarized domains in light beams using analytical and numerical approaches. In particular, our method relies on suitable superpositions of cylindrically symmetric vortex beams with {different} OAM{states $(l_x \neq l_y)$}, distinct spatial parameters, and orthogonal polarization. Along propagation, we observed the formation of rotating chiral domains as a consequence of the superposition. We observed that the number of chiral domains formed is directly related to the topological charges of both components, which can be calculated using the formula $N=|l_x|+|l_y|$ {for} $l_x \cdot l_y<0$ {and $N=||l_x|-|l_y||$ for $l_x \cdot l_y>0$}. Moreover, the phase difference between the two components influences the longitudinal and radial evolution of the domains along propagation. Our results may have implications for various applications in optics namely polarization engineering, laser machining, nanoscale imaging, polarimetry, and light-chiral matter interactions {\cite{Ayuso,KAForbes,Lininger}}.

\section*{Acknowledgements}
This project was funded by the European Research Council (ERC-Consolidator) under the grant agreement No. 101045135 (Beyond Anderson).


\begin{thebibliography}{0}%
\makeatletter
\providecommand \@ifxundefined [1]{%
 \@ifx{#1\undefined}
}%
\providecommand \@ifnum [1]{%
 \ifnum #1\expandafter \@firstoftwo
 \else \expandafter \@secondoftwo
 \fi
}%
\providecommand \@ifx [1]{%
 \ifx #1\expandafter \@firstoftwo
 \else \expandafter \@secondoftwo
 \fi
}%
\providecommand \natexlab [1]{#1}%
\providecommand \enquote  [1]{``#1''}%
\providecommand \bibnamefont  [1]{#1}%
\providecommand \bibfnamefont [1]{#1}%
\providecommand \citenamefont [1]{#1}%
\providecommand \href@noop [0]{\@secondoftwo}%
\providecommand \href [0]{\begingroup \@sanitize@url \@href}%
\providecommand \@href[1]{\@@startlink{#1}\@@href}%
\providecommand \@@href[1]{\endgroup#1\@@endlink}%
\providecommand \@sanitize@url [0]{\catcode `\\12\catcode `\$12\catcode `\&12\catcode `\#12\catcode `\^12\catcode `\_12\catcode `\%12\relax}%
\providecommand \@@startlink[1]{}%
\providecommand \@@endlink[0]{}%
\providecommand \url  [0]{\begingroup\@sanitize@url \@url }%
\providecommand \@url [1]{\endgroup\@href {#1}{\urlprefix }}%
\providecommand \urlprefix  [0]{URL }%
\providecommand \Eprint [0]{\href }%
\providecommand \doibase [0]{https://doi.org/}%
\providecommand \selectlanguage [0]{\@gobble}%
\providecommand \bibinfo  [0]{\@secondoftwo}%
\providecommand \bibfield  [0]{\@secondoftwo}%
\providecommand \translation [1]{[#1]}%
\providecommand \BibitemOpen [0]{}%
\providecommand \bibitemStop [0]{}%
\providecommand \bibitemNoStop [0]{.\EOS\space}%
\providecommand \EOS [0]{\spacefactor3000\relax}%
\providecommand \BibitemShut  [1]{\csname bibitem#1\endcsname}%
\let\auto@bib@innerbib\@empty
%</preamble>
\end{thebibliography}%


\noindent





\begin{thebibliography}{00}

%Structured light.
 \bibitem{Forbes2021} A. Forbes, M. De Oliveira, and M. R. Dennis, "Structured light,"  Nat. Photonics \textbf{15}, 253 (2021). 
https://doi.org/10.1038/s41566-021-00780-4
%Towards higher-dimensional structured light
\bibitem{He2022} C. He, Y. Shen, and  A. Forbes, "Towards higher-dimensional structured light," Light Sci. Appl. \textbf{11}, 205 (2022). https://doi.org/10.1038/s41377-022-00897-3
%Review of structured light in diffuse optical imaging
\bibitem{Angelo2018} J. P. Angelo, Sez-Jade K. Chen, M.n Ochoa, U. Sunar, S. Gioux, and Xavier Intes, "Review of structured light in diffuse optical imaging," J. Biomed. Opt. \textbf{24}, 071602 (2018). https://doi.org/10.1117/1.JBO.24.7.071602
%Structured-Light-Based System for Shape Measurement of the Human Body in Motion
\bibitem{Liberadzki2018} P. Liberadzki, M. Adamczyk, M. Witkowski, and R. Sitnik, "Structured-Light-Based System for Shape Measurement of the Human Body in Motion," Sensors \textbf{18}, 2827 (2018). https://doi.org/10.3390/s18092827
%Faster, sharper, and deeper: structured illumination microscopy for biological imaging
\bibitem{Wu2018} Y. Wu and H. Shroff, "Faster, sharper, and deeper: structured illumination microscopy for biological imaging," Nat. Methods  \textbf{15}, 1011 (2018). https://doi.org/10.1038/s41592-018-0211-z
%Terabit free-space data transmission employing orbital angular momentum multiplexing
\bibitem{Wang2012} J. Wang, Jeng-Yuan Yang, I. M. Fazal, N. Ahmed, Y. Yan, H. Huang, Y. Ren, Y. Yue, S. Dolinar, M. Tur, and A. E. Willner,  "Terabit free-space data transmission employing orbital angular momentum multiplexing," Nat. Photon. \textbf{6}, 488 (2012). https://doi.org/10.1038/nphoton.2012.138
%Terabit-Scale Orbital Angular Momentum Mode Division Multiplexing in Fibers
\bibitem{Bozinovic2013} N. Bozinovic, Y. Yue, Y.Ren, M. Tur, P. Kristensen, H. Huang, A. E Willner, and S. Ramachandran, "Terabit-Scale Orbital Angular Momentum Mode Division Multiplexing in Fibers," Science \textbf{340}, 1545 (2013).  https://doi.org/10.1126/science.1237861

%%%%%%%%%%%%%%%%%%%%%%%%%%%%%%%%%%%%%%%%%%%%%%% POLARIZATION %%%%%%%%%%%%%%%%%%%%%%%%%%%%%%%%%%%%%%%%%%%%%%%
%A new view on polarization microscopy.
\bibitem{Oldenbourg1996} R. Oldenbourg, "A new view on polarization microscopy," Nature \textbf{381}, 811 (1996). https://doi.org/10.1038/381811a0
%A review of complex vector light fields and their applications.
\bibitem{Rosales-Guzman2018} C. Rosales-Guzmán, B. Ndagano, and A. Forbes, "A review of complex vector light fields and their applications," J. Opt. \textbf{20}, 123001 (2018). https://doi.org/10.1088/2040-8986/aaeb7d
%Vectorial light–matter interaction: Exploring spatially structured complex light fields
\bibitem{Wang2020}  J. Wang,  F. Castellucci, and  S. Franke-Arnold, "Vectorial light–matter interaction: Exploring spatially structured complex light fields," AVS Quantum Sci.  \textbf{2}, 031702 (2020). https://doi.org/10.1116/5.0016007
%Polarisation optics for biomedical and clinical applications: a review
\bibitem{He2021} C. He, H. He, J. Chang, B. Chen, H. Ma, and M. J. Booth, "Polarisation optics for biomedical and clinical applications: a review,"  Light Sci. Appl.  \textbf{10}, 194 (2021). https://doi.org/10.1038/s41377-021-00639-x 


\bibitem{Born1999} M. Born and E. Wolf, "Principles of Optics: Electromagnetic Theory of Propagation, Interference ", (7th Edition) (Cambridge University Press, 1999).

\bibitem{Allen1992} L. Allen, M. W. Beijersbergen, R. J. C. Spreeuw, and J. P. Woerdman, "Orbital angular momentum of light and the transformation of Laguerre-Gaussian laser modes," Phys. Rev. A \textbf{45}, 8185 (1992).  https://doi.org/10.1103/PhysRevA.45.8185

{\bibitem{Fickler} R. Fickler, G. Campbell B. Buchler, P. K. Lam,  and A. Zeilinger, "Quantum entanglement of angular momentum states with quantum numbers up to 10,010," Proc. Natl. Acad. Sci. U.S.A. \textbf{113}, 13642 (2016). https://doi.org/10.1073/pnas.1616889113}

\bibitem{Paterson2001} L. Paterson, M. P. MacDonald, J. Arlt, W. Sibbett, and K. Dholakia, "Controlled Rotation of Optically Trapped Microscopic Particles", Science \textbf{292}, 912 (2001).  https://doi.org/10.1126/science.1058591
%%Grier reviewed the optical tweezers and spanner using vortex beams.
\bibitem{Grier2003} D. G. Grier, "A revolution in optical manipulation,"  Nature \textbf{424}, 810 (2003). https://doi.org/10.1038/nature01935
%Barreiro et al. used  OAM in quantum communication beating the capacity limit.
\bibitem{Barreiro2008} J. T. Barreiro, T. C. Wei, and P. G. Kwiat, "Beating the channel capacity limit for linear photonic superdense coding," Nat. Phys. \textbf{4}, 282 (2008).  https://doi.org/10.1038/nphys919

\bibitem{Durnin1987}  J. Durnin, "Exact solutions for nondiffracting beams. I. The scalar theory," J. Opt. Soc. Am. A \textbf{4}, 651 (1987). https://doi.org/10.1364/JOSAA.4.000651
\bibitem{Durnin2005} D. McGloin and K. Dholakia, "Bessel beams: Diffraction in a new light," Contemp. Phys. \textbf{46}, 15 (2005). https://doi.org/10.1080/0010751042000275259

{\bibitem{Gori}F. Gori, G. Guattari, and C. Padovani, “Bessel-Gauss beams,” Opt. Commun. \textbf{64}, 491 (1987). https://doi.org/10.1016/0030-4018(87)90276-8}

\bibitem{Siegman1986} A. E. Siegman, Lasers (University Science Books, 1986).
\bibitem{Paufler2019} W. Paufler, B. Böning, and S. Fritzsche, "High harmonic generation with Laguerre–Gaussian beams," J. Opt \textbf{21}, 094001 (2019). https://doi.org/10.1088/2040-8986/ab31c3

% quasi-diffraction-free Airy beams 1D + 2D
\bibitem{Siviloglou2007} G. A. Siviloglou and D. N. Christodoulides, "Accelerating finite energy Airy beams," Opt. Lett. \textbf{32}, 979 (2007). https://doi.org/10.1364/OL.32.000979
%Observation of Accelerating Airy Beams 1D + 2D
\bibitem{Siviloglou2007exp} G. A. Siviloglou, J. Broky, A. Dogariu, and D. N. Christodoulides, "Observation of Accelerating Airy Beams" Phys. Rev. Lett. \textbf{99}, 213901 (2007). https://doi.org/10.1103/PhysRevLett.99.213901

\bibitem{Efremidis2010} N. K. Efremidis and D. N. Christodoulides, "Abruptly autofocusing waves" Opt. Lett. \textbf{35}, 4045 (2010). https://doi.org/10.1364/OL.35.004045
\bibitem{Papazoglou2011} D. G. Papazoglou, N. K. Efremidis, D. N. Christodoulides, and S. Tzortzakis, "Observation of abruptly autofocusing waves," Opt. Lett. \textbf{36}, 1842 (2011). https://doi.org/10.1364/OL.36.001842
%Accelerating beams review
\bibitem{Efremidis2019} N. K. Efremidis, Z. Chen, M. Segev, and D. N. Christodoulides, "Airy beams and accelerating waves: an overview of recent advances," Optica \textbf{6}, 686 (2019). https://doi.org/10.1364/OPTICA.6.000686

%General for rotating domains
\bibitem{Schechner1996} Y. Y. Schechner, R. Piestun, and J. Shamir, "Wave propagation with rotating intensity distributions," Phys. Rev. E \textbf{54}, R50 (1996). https://doi.org/10.1103/PhysRevE.54.R50
\bibitem{Bekshaev2006} A. Bekshaev and M. Soskin, "Rotational transformations and transverse energy flow in paraxial light beams: linear azimuthons," Opt. Lett. \textbf{31}, 2199 (2006). https://doi.org/10.1364/OL.31.002199
\bibitem{Bekshaev2006_2} A. Ya. Bekshaev M. S. Soskin and M. V. Vasnetsov, "Centrifugal transformation of the transverse structure of freely propagating paraxial light beams," Opt. Lett. \textbf{31}, 694 (2006). https://doi.org/10.1364/OL.31.000694

\bibitem{McGloin2003} D. McGloin, V. Garces-Chavez, and K. Dholakia, "Interfering Bessel beams for optical micromanipulation," Opt. Lett. \textbf{28}, 657 (2003). https://doi.org/10.1364/OL.28.000657
\bibitem{Vasilyeu2009} R. Vasilyeu, A. Dudley, N. Khilo, and A. Forbes, "Generating superpositions of higher–order Bessel beams," Opt. Express \textbf{17}, 23389 (2009). https://doi.org/10.1364/OE.17.023389
\bibitem{Rop2012} R. Rop, A. Dudley, C. Lopez-Mariscal, and A. Forbes, "Measuring the rotation rates of superpositions of higher-order Bessel beams," J. Mod. Opt. \textbf{59}, 259 (2012). https://doi.org/10.1080/09500340.2011.631714
\bibitem{Schulze2015} C. Schulze, F. S. Roux, A. Dudley, R. Rop, M. Duparre, and A. Forbes, "Accelerated rotation with orbital angular momentum modes," Phys. Rev. A \textbf{91}, 043821 (2015). https://doi.org/10.1103/PhysRevA.91.043821

\bibitem{Webster2017} J. Webster, C. Rosales-Guzman, and A. Forbes, "Radially dependent angular acceleration of twisted light," Opt. Lett. \textbf{42}, 675 (2017). https://doi.org/10.1364/OL.42.000675
%Broaden family of ToWs 
%Tornado waves
\bibitem{Brimis2020} A. Brimis, K. G. Makris, and D. G. Papazoglou, "Tornado waves," Opt. Lett. \textbf{45}, 280 (2020). https://doi.org/10.1364/OL.45.000280
%Spiraling light: Generating optical tornados
\bibitem{Mansour2022} D. Mansour, A. Brimis, K. G. Makris, and D. G. Papazoglou, "Spiraling light: Generating optical tornados," Phys. Rev. A \textbf{105}, 053514 (2022). https://doi.org/10.1103/PhysRevA.105.053514

\bibitem{Brimis2023} A. Brimis, K. G. Makris, and D. G. Papazoglou, "Optical vortices shape optical tornados," Opt. Express \textbf{31}, 27582 (2023).
https://doi.org/10.1364/OE.495836


%L-lines
\bibitem{Nye_book}  J.F. Nye, "Natural Focusing and the Fine Structure of Light", (IOP Bristol 1999).
%Elliptic critical points in paraxial optical fields
\bibitem{Freund2002b} I. Freund, M. Soskin, and A. Mokhun, "Elliptic critical points in paraxial optical fields," Opt. Commun. \textbf{208}, 223 (2002). https://doi.org/10.1016/S0030-4018(02)01585-7


%Full Poincaré beams 
\bibitem{Beckley2010} A. M. Beckley, T. G. Brown, and M. A. Alonso, "Full Poincaré beams," Opt. Express \textbf{18}, 10777 (2010). https://doi.org/10.1364/OE.18.010777

%On the overall polarisation properties of Poincaré beams
\bibitem{Lopez-Mago2019} D. Lopez-Mago, "On the overall polarisation properties of Poincaré beams," J. Opt. \textbf{21}, 115605 (2019).  https://doi.org/10.1088/2040-8986/ab4c25 
%Poincaré Bessel beams: structure and propagation
\bibitem{Holmes2019} B. M. Holmes and E. J. Galvez, "Poincaré Bessel beams: structure and propagation" J. Opt. \textbf{21}, 104001 (2019).  https://doi.org/10.1088/2040-8986/ab3d7d
%Full Poincaré beam with all the Stokes vortices
\bibitem{Arora2019} G. Arora, Ruchi, and P. Senthilkumaran, "Full Poincaré beam with all the Stokes vortices," Opt. Lett. \textbf{44}, 5638 (2019). https://doi.org/10.1364/OL.44.005638
%Optical currents in Poincaré beams
\bibitem{Ruchi2020} Ruchi, and P. Senthilkumaran, "Optical currents in Poincaré beams," Phys. Rev. A \textbf{102}, 013509 (2020). https://doi.org/10.1103/PhysRevA.102.013509
%Shaping light in 3d space by counter-propagation
\bibitem{Droop2021}R. Droop, E. Asché, E. Otte, and C. Denz, "Shaping light in 3d space by counter-propagation," Sci. Rep. \textbf{11}, 18019 (2021). https://doi.org/10.1038/s41598-021-97313-4 
%Higher-Order Poincaré´ Sphere, Stokes Parameters, and the Angular Momentum of Light
\bibitem{Milione2011} G. Milione, H. I. Sztul, D. A. Nolan, and R. R. Alfano, "Higher-Order Poincaré Sphere, Stokes Parameters, and the Angular Momentum of Light," Phys. Rev. Lett. \textbf{107}, 053601 (2011). https://doi.org/10.1103/PhysRevLett.107.053601 
%Poincaré-beam patterns produced by nonseparable superpositions of Laguerre–Gauss and polarization modes of light
\bibitem{Galvez2012} E. J. Galvez, S. Khadka, W. H. Schubert, and S. Nomoto, "Poincaré-beam patterns produced by nonseparable superpositions of Laguerre–Gauss and polarization modes of light," Appl. Opt. \textbf{51}, 2925 (2012). https://doi.org/10.1364/AO.51.002925
%Dynamic modulation of Poincaré beams
\bibitem{Alpmann2017} C. Alpmann, C. Schlickriede, E. Otte, and C. Denz, "Dynamic modulation of Poincaré beams," Sci. Rep. \textbf{7}, 8076 (2017). https://doi.org/10.1038/s41598-017-07437-9

% Chapter 5 Singular Optics: Optical Vortices and Polarization Singularities
\bibitem{Dennis2009} M. R. Dennis, K. O’Holleran, and M. J. Padgett, "Chapter 5 Singular Optics: Optical Vortices and Polarization Singularities," Prog. Opt. \textbf{53}, 293 (2009). https://doi.org/10.1016/S0079-6638(08)00205-9
% Visualizing polarization singularities in Bessel–Poincaré beams 
 \bibitem{Shvedov2015} V. Shvedov, P. Karpinski, Y. Sheng, X. Chen, W. Zhu, W. Krolikowski, and C. Hnatovsky, "Visualizing polarization singularities in Bessel–Poincaré beams," Opt. Express \textbf{23}, 12444 (2015). https://doi.org/10.1364/OE.23.012444

%Optical forces on submicron particles induced by full Poincaré beams
\bibitem{Wang3} Li-Gang Wang, "Optical forces on submicron particles induced by full Poincaré beams," Opt. Express \textbf{20}, 20814 (2012). https://doi.org/10.1364/OE.20.020814

%%%% Rotating optical activity using Poincare beams %%%%
%Gouy phase effects on propagation of pure and hybrid vector beams
\bibitem{Sanchez-Lopez2019} M. M. Sánchez-López, J. A. Davis, I. Moreno, A. Cofré, and D. M. Cottrell, "Gouy phase effects on propagation of pure and hybrid vector beams," Opt. Express \textbf{27}, 2374 (2019). https://doi.org/10.1364/OE.27.002374
%Self-accelerated optical activity in free space induced by the Gouy phase
\bibitem{Li2020} P. Li, X. Fan, D. Wu, S.Liu, Y. Li, and J. Zhao, "Self-accelerated optical activity in free space induced by the Gouy phase," Photon. Res. \textbf{8}, 475 (2020). https://doi.org/10.1364/PRJ.380675
\bibitem{Li2018} P. Li, D. Wu, Y. Zhang, S. Liu, Y. Li, S. Qi, and J. Zhao, "Polarization oscillating beams constructed by copropagating optical frozen waves," Phot. Research  \textbf{6}, 756 (2018). https://doi.org/10.1364/PRJ.6.000756


%Accelerating polarization structures in vectorial fields
\bibitem{Singh2021} K. Singh, W. T. Buono, A. Forbes, and A. Dudley, "Accelerating polarization structures in vectorial fields," Opt. Express \textbf{29}, 2727 (2021). https://doi.org/10.1364/OE.411029
%Rotating full Poincaré beams
\bibitem{Krasnoshchekov2017} Ye. A. Krasnoshchekov, V. V. Yaparov, and V. B. Taranenko, "Rotating full Poincaré beams," Ukr. J. Phys. Opt. \textbf{18}, 1 (2017). 
%Controlled generation of rotating vector beams in quasi-isotropic laser
\bibitem{Krasnoshchekov2019} Ye. A. Krasnoshchekov, V. V. Yaparov, and V. B. Taranenko, "Controlled generation of rotating vector beams in quasi-isotropic laser," Ukr. J. Phys. Opt. \textbf{20}, 1 (2019). 


%Ring-Airy and vortex (added)
\bibitem{Jiang2012} Y. Jiang, K. Huang, and X. Lu, "Propagation dynamics of abruptly autofocusing Airy beams with optical vortices," Opt. Express \textbf{20}, 18579 (2012). https://doi.org/10.1364/OE.20.018579
%LG modes
\bibitem{Kotlyar1997} V. V. Kotlyar, V. A. Soifer, and S. N. Khonina, "Rotation of multimode Gauss-Laguerre light beams in free space," Tech. Phys. Lett. \textbf{23}, 657 (1997). https://doi.org/10.1134/1.1261648
\bibitem{Kotlyar2007} V. V. Kotlyar, S. N. Khonina, R. V. Skidanov, and V. A. Soifer, "Rotation of laser beams with zero of the orbital angular momentum," Opt. Commun. \textbf{274}, 8 (2007). https://doi.org/10.1016/j.optcom.2007.01.059
%%%%%%% Broaden family of ToWs %%%%%%%
%Abruptly autofocusing polycyclic tornado ring Airy beam
\bibitem{Wu2021} Y. Wu, C. Xu, Z. Lin, H. Qiu, X. Fu, K. Chen, and D. Deng, "Abruptly autofocusing polycyclic tornado ring Airy beam," New J. Phys. \textbf{22}, 093045 (2021). https://doi.org/10.1088/1367-2630/abb125
%Generation and control of tornado waves by means of ring swallowtail vortex beams
\bibitem{Jiang2022} J. Jiang, D. Xu, Z. Mo, X. Cai, H. Huang, Y. Zhang, H. Yang, H. Huang, Y. Wu, L. Shui, and D. Deng, "Generation and control of tornado waves by means of ring swallowtail vortex beams," Opt. Express \textbf{30}, 11331 (2022). https://doi.org/10.1364/OE.453165
%Experimental generation of the polycyclic tornado circular swallowtail beam with self-healing and auto-focusing
\bibitem{Zhang2022}Y. Zhang, J. Tu, S. He, Y. Ding, Z. Lu, Y. Wu, G. Wang, X. Yang, and D. Deng, "Experimental generation of the polycyclic tornado circular swallowtail beam with self-healing and auto-focusing," Opt. Express \textbf{30}, 1829 (2022). https://doi.org/10.1364/OE.446818
%Switchable optical ring lattice in free space
\bibitem{Xu2023} D. Xu, T. Qi, Y. Chen, and W. Gao, "Switchable optical ring lattice in free space," Opt. Express \textbf{31}, 9416 (2023). https://doi.org/10.1364/OE.485612

\bibitem{Panagiotopoulos2013} P. Panagiotopoulos, D. Papazoglou, A. Couairon, and S. Tzortzakis, "Sharply autofocused ring-Airy beams transforming into non-linear intense light bullets," Nat. Commun. \textbf{4},  2622 (2013). https://doi.org/10.1038/ncomms3622

{\bibitem{Ayuso} D. Ayuso, O. Neufeld, A. F. Ordonez, P. Decleva, G. Lerner, O. Cohen, M. Ivanov, and O. Smirnova, “Synthetic chiral light for efficient control of chiral light–matter interaction,” Nat. Photonics \textbf{13}, 866 (2019). https://doi.org/10.1038/s41566-019-0531-2}
{\bibitem{KAForbes} K. A. Forbes and D. L. Andrews, “Orbital angular momentum of twisted light: chirality and optical activity,” J. Phys. Photonics \textbf{3}, 022007 (2021). https://doi.org/10.1088/2515-7647/abdb06}
{\bibitem{Lininger} A. Lininger, G. Palermo, A. Guglielmelli, G. Nicoletta, M. Goel, M. Hinczewski, and G. Strangi, “Chirality in Light–Matter Interaction,” Adv. Mater., 2107325 (2022). https://doi.org/10.1002/adma.202107325}

\end{thebibliography}
\end{document}